\documentclass{article}
\usepackage{ijcai24}

\usepackage{times}
\usepackage{soul}
\usepackage{url}
\usepackage[hidelinks]{hyperref}
\usepackage[utf8]{inputenc}
\usepackage[small]{caption}
\usepackage{graphicx}
\usepackage{amsmath}
\usepackage{amsthm}
\usepackage{booktabs}
\usepackage{algorithm}
\usepackage{algorithmic}
\usepackage[switch]{lineno}

\title{Music Consistency Models}

\author{
    Zhengcong Fei, Mingyuan Fan, Junshi Huang
    \affiliations
    Kunlun Inc.
    \emails
    \{feizhengcong\}@gmail.com
}

\begin{document}

\maketitle

\begin{abstract}
    Consistency models have exhibited remarkable capabilities in facilitating efficient image/video generation, enabling synthesis with minimal sampling steps. It has proven to be advantageous in mitigating the computational burdens associated with diffusion models. Nevertheless, the application of consistency models in music generation remains largely unexplored. To address this gap, we present Music Consistency Models (\texttt{MusicCM}), which leverages the concept of consistency models to efficiently synthesize mel-spectrogram for music clips, maintaining high quality while minimizing the number of sampling steps. Building upon existing text-to-music diffusion models, the \texttt{MusicCM} model incorporates consistency distillation and adversarial discriminator training. Moreover, we find it beneficial to generate extended coherent music by incorporating multiple diffusion processes with shared constraints. Experimental results reveal the effectiveness of our model in terms of computational efficiency, fidelity, and naturalness. Notable, \texttt{MusicCM} achieves seamless music synthesis with a mere four sampling steps, e.g., only one second per minute of the music clip, showcasing the potential for real-time application. 
\end{abstract}

\section{Introduction}

In recent years, the field of text-to-music generation has witnessed tremendous progress in synthesizing natural and coherent music clips, primarily driven by the development of diffusion models. 
Several diffusion-based methods such as Noise2Music \cite{huang2023noise2music}, Mousai \cite{schneider2023mo}, MusicLDM \cite{chen2023musicldm}, Make-An-Audio \cite{huang2023make}, and AudioLDM \cite{liu2023audioldm2,liu2023audioldm,zhu2023survey}, have achieved notable performance by integrating additional audio mel-spectrogram into existing image diffusion models \cite{rombach2022high,ramesh2022hierarchical} to effectively handle the temporal and spectral characteristics. 
However, these diffusion-based approaches inherently necessitate a considerable number of sampling steps during the music synthesis process in inference, e.g., 50-step DDIM sampling \cite{song2020denoising}. 
This limitation poses an obstacle to the efficient and expeditious generation of high-quality musical compositions.

To address the issue of high sampling cost in diffusion models, the concept of consistency models has been introduced within the fields of image/video generation \cite{song2023consistency,song2023improved,luo2023latent,luo2023lcm,xiao2023ccm,wang2023videolcm}. 
These methods are designed to generate samples in a single step while retaining the crucial advantages of diffusion models. These advantages include the ability to trade computational resources for sample quality through multi-step sampling and the capability of performing zero-shot data editing \cite{yang2023diffusion}. By enabling efficient synthesis with a minimal number of steps, such models have achieved remarkable progress, reducing the required steps from 50 to as few as 4. Despite the notable success achieved in image generation, the application of consistency models in the realm of music synthesis remains largely unexplored.

Following this premise, we present the Music Consistency Models, denoted as \texttt{MusicCM}, a diffusion-based consistency model designed for music creation according to text prompts. 
Our approach is centered around the utilization of existing diffusion models within the domain of music generation, while also incorporating the concepts of consistency distillation. 
Specifically, by implementing the \texttt{MusicCM} framework, our objective is to mitigate the necessity for extensive sampling procedures while concurrently upholding the production of high-quality synthesized music. Through the utilization of adversarial discrimination \cite{sauer2023adversarial,fei2020actor}, the model is compelled to generate samples that reside directly on the manifold of authentic musical compositions during each forward pass, thereby circumventing issues such as blurriness and other artifacts often encountered in alternative distillation methods.
In light of the works \cite{bar2023multidiffusion,wang2023unlimited,fei2021memory}, we further introduce an enhanced generation process that combines several reference diffusion generation processes bound together with a set of shared constraints. In between, \texttt{MusicCM} is applied to different regions in the generated mel-spectrogram, serving as a reference, predicting a denoising sampling step for each. Subsequently, a global denoising sampling step is taken that harmonizes these individual steps through a least squares optimal solution, ultimately resulting in a unified and coherent synthesis.

Experimentally, we have obtained quantitative and qualitative findings that unequivocally validate the efficacy of our proposed approach. Notably, by integrating the aforementioned techniques, our method attains a remarkable level of fidelity in music synthesis, employing a mere 4 to 6 sampling steps, equivalent to approximately one second for each minute of music clips. This outcome serves as a testament to the potential of our method for facilitating rapid and real-time synthesis processes.

\textbf{Contributions.} To sum up, our contributions are as follows: 
($\textbf{i}$) We present MusicCM, a novel framework that aims to bridge the gap between diffusion models and consistency models in the domain of music generation. Our approach incorporates consistency distillation and an adversarial discriminator to enable the efficient synthesis of high-quality music clips. 
($\textbf{ii}$) To address the challenges of maintaining long music consistency while ensuring memory efficiency, we introduce multiple diffusion processes with shared constraints during the inference stage. 
($\textbf{iii}$) We provide both quantitative and qualitative results to demonstrate the effectiveness of our MusicCM approach. 
To improve reproducibility, we will publicly release the source code and trained models of all experiments. 
Finally, by exploring the potential of consistency models in music generation, we aim to contribute to the field of fast music synthesis and provide a simplified and effective baseline for future research. 

\section{Related Works}


\paragraph{Text-to-image generation.}
Remarkable advancements have been achieved in text-to-image synthesis, primarily attributed to the emergence of generative models 
\cite{goodfellow2014generative,ho2022classifier,kang2023scaling,shen2021closed,fei2022progressive}. Especially diffusion models \cite{kawar2023imagic,mou2023t2i,nichol2021glide,rombach2022high,ruiz2023dreambooth} play a crucial role. 
Various methodologies, such as \cite{ramesh2022hierarchical,gafni2022make}, propose a two-stage approach where the input text undergoes conversion into image embeddings using a prior model, which is then utilized for image synthesis. Stable Diffusion \cite{rombach2022high}
introduces a VAE-based approach in the latent space to decrease
computational demand and optimizes the model with
large-scale datasets \cite{schuhmann2022laion}. 
\cite{sauer2023adversarial} use score distillation from a teacher models combination with an adversarial loss to maintain high quality with few-step settings. 
Subsequent methods \cite{zhang2023adding,fei2023gradient,fei2024diffusion,fei2024scalable,huang2023composer} have incorporated additional conditional inputs, such as depth maps or sketches, for spatially controllable image synthesis.

\paragraph{Consistency model.}
Inference cost is an important factors for multimodal generation \cite{fei2019fast,fei2021partially,yan2021semi,fei2022deecap}.
Consistency model \cite{song2023consistency} has been developed to address the limitation of numerous inference steps in diffusion models. By leveraging the probability flow ordinary differential equation, consistency models aim to learn the mapping of any point at any given time step to the initial point of the trajectory, which corresponds to the original clean image. 
The introduction of the consistency model allows for efficient one-step image generation without compromising the benefits of multi-step iterative sampling.  Consequently, this approach also facilitates the production of high-quality results through multi-step inference.  Moreover, LCM \cite{luo2023latent} explores consistency models in the latent space to save memory consumption and improve inference efficiency. Subsequently, several methods \cite{luo2023lcm,sauer2023adversarial,xiao2023ccm,wang2023videolcm} have also delved into efficient generation techniques, building upon the foundation set by the consistency model, and achieved remarkable outcomes. Inspired by the success of the consistency model in the realm of image/video generation, we propose extending its application to the domain of music generation.

\begin{figure*}[t]
   \centering
   \includegraphics[width=0.95\linewidth]{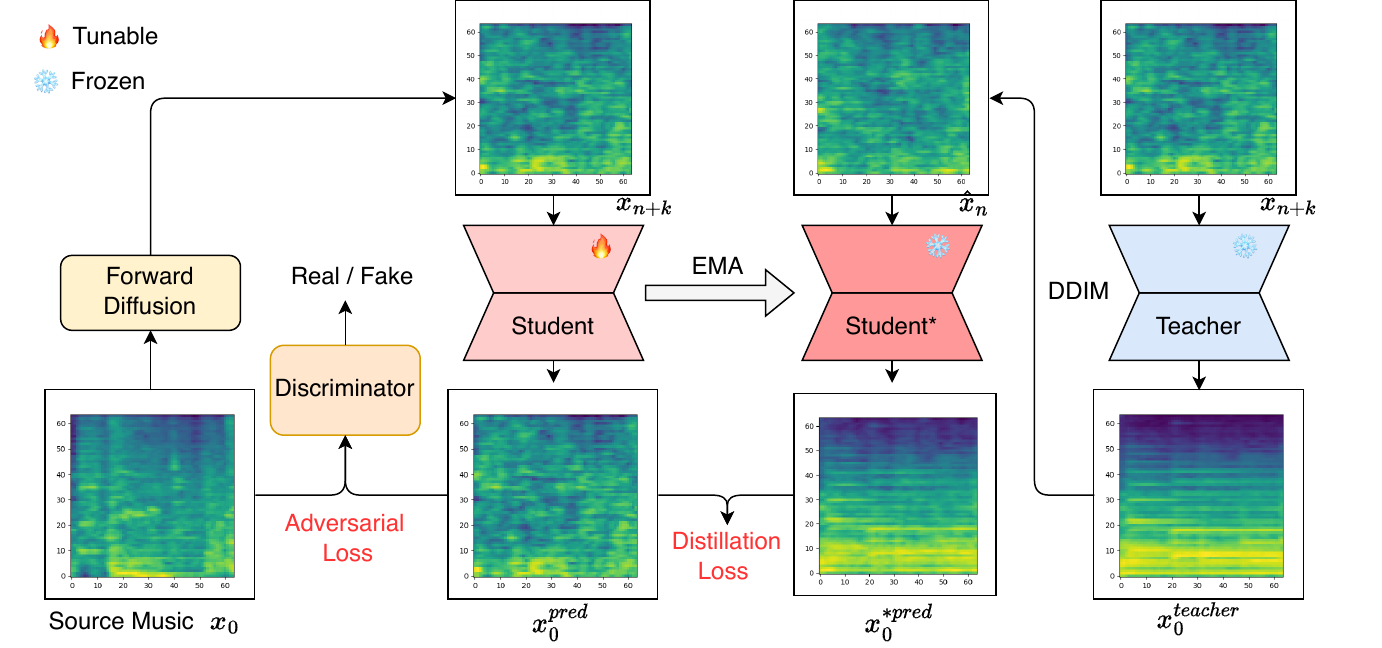}
   \caption{\textbf{Overview of music consistency models.}  Given a source music mel-spectrogram $x_0$, a forward diffusion operation is first performed to add noise to the music. Then, the noised $x_{n+k}$ is entered into the student and teacher model to predict music clips. $\hat{x}_n$ is estimated by the teacher model and fed into the EMA student model. To learn self-consistency, a distillation loss is imposed to constrain the output of the two student models to be consistent, and an adversarial loss is used to fool a discriminator which is trained to distinguish the generated samples $x_0^{pred}$ from real music $x_0$. The whole consistency distillation is conducted in the latent space, and conditional guidance is omitted for ease of presentation. The teacher model is a music diffusion model, and the student shares the same network structure as the teacher model and is initialized with the parameters of the teacher model. 
   }
   \label{fig:1}
\end{figure*}

\paragraph{Conditional music generation.}

Numerous studies have been conducted on the audio generation guided by text, including Diffsound \cite{yang2023diffsound}, AudioGen \cite{kreuk2022audiogen}, AudioLDM \cite{audioldm-liu2023audioldm}, and Make-an-Audio \cite{make-an-audio-huang2023make} showing impressive results. 
In the domain of music, there exist text-to-music models such as retrieval-based MuBERT \cite{MubertAI}, language-model-based MusicLM \cite{agostinelli2023musiclm}, diffusion-based Riffusion \cite{Forsgren_Martiros_2022} Noise2Music \cite{huang2023noise2music} and so on \cite{li2023jen,melechovsky2023mustango}. 
However, music models often necessitate substantial quantities of privately owned music data that are inaccessible to the public, impeding the reproducibility and further development of research efforts. 
Among these aforementioned models, MusicLDM is based on open-source Stable Diffusion \cite{stable-diffusion-rombach2022high}, CLAP \cite{clap-wu2023large,fei2023jepa}, and HiFi-GAN \cite{kong2020hifi} architectures. Therefore, we base our approach on these, to create \texttt{MusicCM} for our experiments.

\section{Methodology}

The proposed \texttt{MusicCM} framework extends the principles of consistency models. Specifically, we begin by providing a concise overview of consistency models. Subsequently, we delve into the intricate details of the proposed \texttt{MusicCM} framework. A presentation of the overall structure is depicted in Figure \ref{fig:1}. Lastly, we devise a restriction mechanism to address the challenges associated with lengthy and coherent music creation.

\subsection{Preliminaries: Consistency Models}

In order to expedite the image generation, \cite{song2023consistency} brings into the conception of the consistency model. It endeavors to enhance a learning framework that can efficiently map any given point in time to the initial point of the PF-ODE trajectory. Formally, the self-consistency property can be formulated as:
\begin{equation}
    f_\theta (x_t, t) = f_\theta(x_{t'}, t'), \forall t, t' \in [\epsilon, T],
\end{equation}
where $\epsilon$ is a time step, $T$ is the overall denoising step, and $x_t$ denotes the noised input.

To accelerate the training and extract the strong prior knowledge of the established diffusion models \cite{rombach2022high}, consistency distillation is proposed as:
\begin{equation}
    L_{distil}(\theta, \theta*; \Phi) = {E}[d(f_\theta(x_{t_{n+1}}, t_{n+1}), f_{\theta*}(\hat{x}_{t_n}, t_n))]
\end{equation}
where $\Phi$ means the applied ODE solver and the model parameters $\theta*$ are obtained from the exponential moving average (EMA) of $\theta$. $\hat{x}_{t_n}$ is the estimation of $x_{t_n}$ according to:
\begin{equation}
    \hat{x}_{t_n} \leftarrow x_{t_{n+1}} + (t_n - t_{n-1}) \Phi (x_{t_{n+1}}, t_{n+1}) \label{eq:3}
\end{equation} 
LCM \cite{luo2023latent} conducts the above consistency optimization in the latent space and applies classifier guidance \cite{ho2022classifier} in Equation \ref{eq:3} to inject control signals, such as textual prompts. For more details, please refer to the original works \cite{song2020denoising}.

\subsection{Application to Music}

The proposed \texttt{MusicCM} is also established within the latent space to mitigate the computational burden, in accordance with LCM. 
To harness the substantial knowledge encapsulated in pre-trained music diffusion models and speed up the training process, we employ consistency distillation and adversarial discrimination strategies. It is pertinent to observe that the pre-trained diffusion models under consideration may encompass diverse typologies, e.g., MusicLDM and Noise2Music. 

In \texttt{MusicCM}, we apply DDIM \cite{song2020denoising} as the basic ODE solver $\Psi$ to estimate $\hat{x}_{t_n}$ as:
\begin{equation}
    \hat{x}_{t_n} \approx x_{t_{n+1}} + \Psi(x_{t_{n+1}}, t_{n+1}, t_n, c)
\end{equation}
where $c$ means the conditional inputs, which can be textual prompts in a text-to-music generation.

\begin{figure*}[t]
   \centering
   \includegraphics[width=0.95\linewidth]{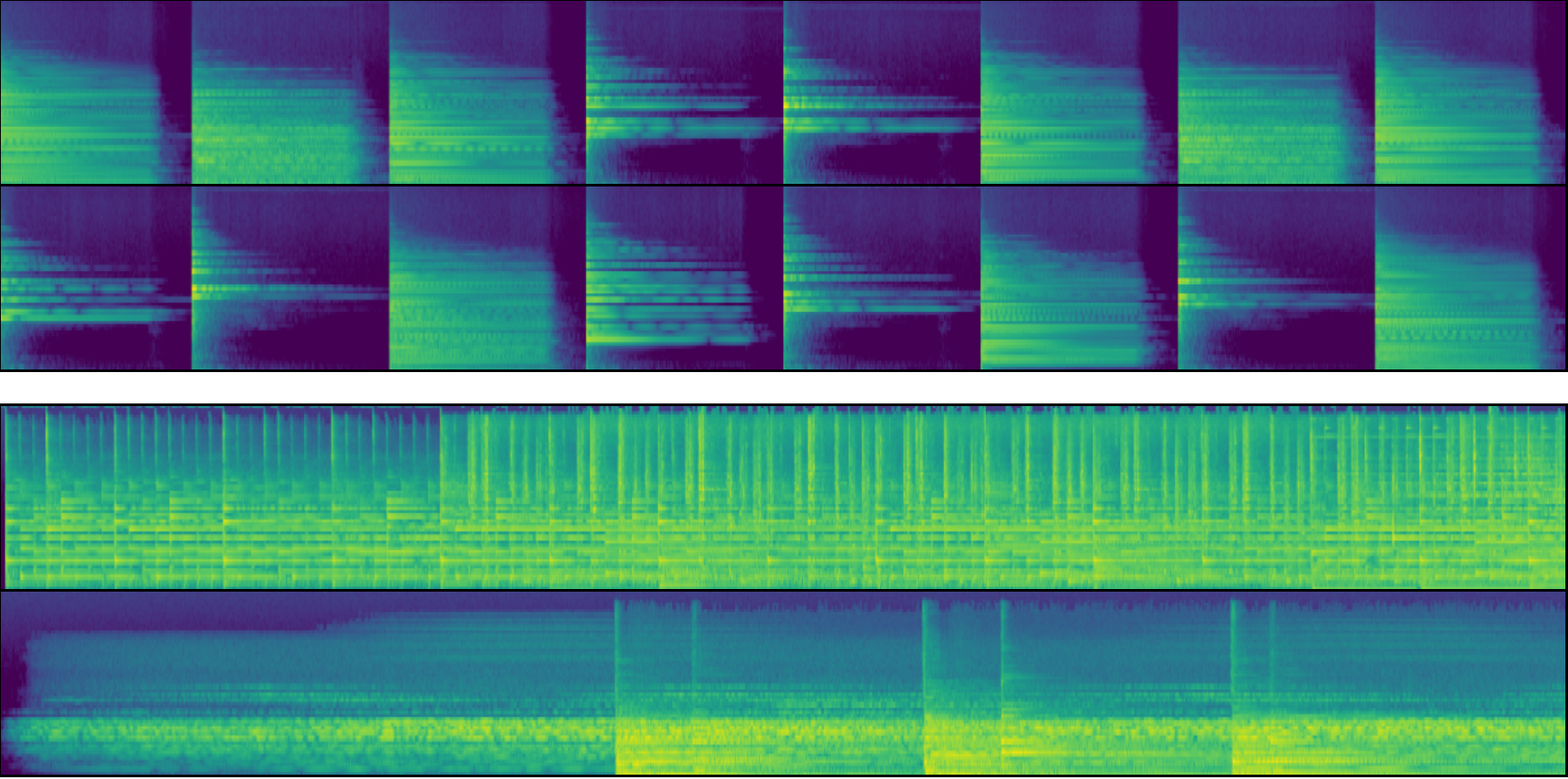}
   \caption{\textbf{Comparison of long music generation through independent paths vs. shared restricted paths.} 
   Input text prompt: \emph{Bright, cheerful and groovy piano, classical}.  
   As expected, there is no coherency between clips in independent; Starting from the same noise, our shared restriction process steers these initial diffusion paths into consistent and high quality music clips. 
   }
   \label{fig:2}
\end{figure*}

\paragraph{Classifier-free guidance.}

Given the pivotal role of classifier-free guidance in the synthesis of high-quality content, we extend its application to the consistency distillation stage, introducing a parameter denoted as $w$ to regulate the scale of guidance: 
\begin{align}
    \hat{x}_{t_n} \approx & x_{t_{n+1}} + (1 + w) \Psi (x_{t_{n+1}}, t_{n+1}, t_n, c) \\
    & -w \Psi(x_{t_{n+1}}, t_{n+1}, t_n, \phi)
\end{align} 
To maintain alignment of initial parameters and design of the consistency model consistent with the teacher diffusion model, we train the consistency model with a fixed value of $w$, such as 2.0.  It is noteworthy that classifier-free guidance is exclusively implemented in the training phase for the teacher diffusion model and is not deemed essential during the inference process of the consistency model. 
During the inference phase, we can sample 4$\sim$6 LCM steps to produce plausible results on text-to-music generation.

\paragraph{Adversarial discriminator.} 

We follow the discriminator design in \cite{sauer2023stylegan}, 
where a frozen pre-trained feature network $F$ and a set of trainable lightweight discriminator head $D_{\varphi, k}$, which are applied on features at different layers of $F$. Similar to \cite{sauer2023adversarial}, the discriminator is conditioned on text embedding and mel-spectrum embedding. 
We then utilize a hinge loss \cite{lim2017geometric} as the adversarial objective function:
\begin{equation}
    L^G_{adv}(\hat{x}_{t_n}, \varphi) = -E[\sum_k D_{\varphi,k}(F_k(\hat{x}_{t_n}))]
\end{equation}
whereas the discriminator is trained to minimize as:
\begin{equation}
    \begin{split}
         & L^D_{adv}(\hat{x}_{t_n}; \varphi)  \\
        &=  E[\sum_k \max(0, 1 - D_{\varphi,k}(F_k(x_0)))+\gamma R1(\varphi)] \\
        & + E[\sum_k \max(0, 1 + D_{\varphi,k}(F_k(\hat{x}_{t_n})))]
    \end{split}
\end{equation}
Where R1 denotes the R1 gradient penalty \cite{mescheder2018training}. 
Rather than computing the gradient penalty with respect to the pixel values, we compute it on the input of each discriminator head $D_{\psi, k}$.
Finally, based on the above analysis, the overall objective is:
\begin{equation}
    L = L_{adv}^G + \lambda L_{distil}
\end{equation}
where $\lambda$ is the balancing factor.

\begin{algorithm}[t]
    \caption{Shared restricted generation for long music}\label{alg:FDS}
    \textbf{Input}{ \texttt{MusicCM} model $\Psi$,  resolution of the desired music $H', W'$, mappings defining crops from the music $\{g_i\}_{i=1}^n$, conditioned text $c$;
    }
    \begin{algorithmic}[1] 
    \STATE $\triangleright$ Noise initialization
    \STATE $X_T^l \sim \mathcal{N}(0, I), X_T^l \in \mathcal{R}^{H' \times W'}$ 
    \FOR{t = T,..., 1} 
    \STATE $\triangleright$ Take crops from mel-spectrograms
    \STATE $X_t^{i,s} \leftarrow g_i(X_t^l)$ 
    \STATE $\triangleright$ Per-crop diffusion updates
    \STATE $X^{i,s}_{t-1} \leftarrow \Psi(X^{i,s}_t, z)$
    \STATE $X_{t-1}^l \leftarrow \pi (X_t^l, z)$
    \ENDFOR
\end{algorithmic}
\end{algorithm}

\begin{table*}[t]
\caption{\textbf{The evaluation of generation quality among MusicLDMs and other text-to-music baselines.} All experiments are performed on a single NVIDIA A100 GPU. The inference overhead of generating 10s music clips at a time is reported as average for each music clip. }
\label{tab:1}
\centering 
\resizebox{0.95\textwidth}{!}{
\begin{tabular}{llccccr}
\toprule
\textbf{Model} & \textbf{Step} & \textbf{FD}$_{pann}$ $\downarrow$ & \textbf{FD}$_{vgg}$ $\downarrow$ & \textbf{Inception Score} $\uparrow$ & \textbf{KL Div.} $\downarrow$ & \textbf{Latency} \\ \midrule 
Riffusion \cite{Forsgren_Martiros_2022}  &50  & 68.95 & 10.77 & 1.34 & 5.00 &- \\
MuBERT \cite{MubertAI} &-  & 31.70 & 19.04 & 1.51 & 4.69 &-\\
AudioLDM \cite{audioldm-liu2023audioldm} &50 & 38.92 & 3.08 & 1.67 & 3.65 & - \\ 
MusicLDM \cite{chen2023musicldm} & 50  & 26.67 & 2.40 & {1.81} & 3.80 & 2.19s\\
\texttt{MusicCM} & 4 & 27.13 &2.48 &1.78 & 3.88 &0.37s \\
\texttt{MusicCM} & 1 & 32.12 &3.31 &1.60  & 4.12 &0.23s \\
 \bottomrule 
\end{tabular}
}
\end{table*}

\subsection{Shared Restricted Process for Long Music}

In light of the memory constraints inherent in the music diffusion model when generating prolonged and coherent musical sequences, we introduce a parallel restriction process, as outlined in \cite{bar2023multidiffusion}. 
Specifically, at each denoising generation step, we amalgamate the denoising directions, supplied by the reference \texttt{MusicCM} model, from all the crops, and strive to follow them all as closely as possible, constrained by the fact that nearby crops share common value in mel-spectrogram.  
Intuitively, we encourage each crop to be a real sample from the reference \texttt{MusicCM} model. It is crucial to note that despite potential variations in denoising directions among individual crops, our framework yields a cohesive denoising step, resulting in the production of high-quality and seamlessly connected long-form music.

Formally, define a long mel-spectrograms space $\mathcal{X}^{l}$ with $H'\geq H$, $W'\geq H$ directly from a trained consistency music model $\Psi$ working in space $\mathcal{X}^{s}$. 
Let $g_i(X^s)\in \mathcal{X}^{s}$ is an $i$-th $H\times W$ crop of image $X^l$, and $z$ serve as condition space mapping from corresponding set. We consider $n$ such crops that cover the original images $X^l$ and get as:
\begin{equation}
\label{eq:panorama}
\small 
    \pi(X_t^l,z)  = \text{argmin}_{X^l \in \mathcal{X}^l}    \sum_{i=1}^n || g_i(X^l) - \Psi(g_i(X^l), z) ||^2 
\end{equation}
that is a least-squares problem, the solution of which can be calculated analytically. 
Mapping function $g_i$ is defined as fixed-size crops from the full mel-spectrogram. 
The entire process is listed in Algorithm \ref{alg:FDS}.
Our maps $g_i,..., g_n$ provide crops with a sliding window of size ${step}$ in the latent space. In particular, $n={\frac{H'-H}{{step}} \cdot {\frac{W'- W}{{step}}}}$. Note that we can compute the per-crop diffusion updates in parallel (i.e., in a batch), resulting in a total of ${\frac{{T \cdot n}}{b}}$ calls to the reference \texttt{MusicCM} $\Psi$, where $b$ denotes the batch size. Note that while each denoising steps $\Psi(g_i(X^l))$ may pull to a different direction, our process fuses these inconsistent directions into a global denoising step $\Psi(x^l)$, resulting in a high-quality seamless music clips.

Taking a more intuitive step, we illustrate the property in Figure \ref{fig:2}, where we consider a long music of $H \times 8W$. We also show results when independently applying $\Psi$ on eight non-overlapping crops. As expected, there is no coherency between crops since this amounts to eight random samples. Starting from the same initial noise, our generation process, allows us to fuse these initially-unrelated diffusion paths, and steer the generation into high-quality, coherent music.

\section{Experiments}

In this section, we commence by elucidating the particulars of the experimental configurations. Subsequently, an assessment of \texttt{MusicCM} ensues, focusing on text-music relevance, novelty, and generation temporal efficiency through established metrics. Lastly, a subjective listening test is undertaken to provide a supplementary evaluation.

\subsection{Experimental Setup}

\paragraph{Datasets.}
Our music consistency model undergoes training utilizing two extensively employed datasets, namely Audiostock \cite{chen2023musicldm} and MagnaTagATune \cite{bogdanov2019mtg}, in conjunction with the frozen VAE and Hifi-GAN. The Audiostock dataset encompasses 9,000 music tracks for the training phase, contributing to a cumulative duration of 455.6 hours. Each track within this dataset is accompanied by an accurate textual description. To augment the diversity of musical styles and facilitate emotional perception, an additional 5,000 music clips are incorporated, with an average sampling approach from MagnaTagATune based on the statistical analysis of the top 50 tags.

\paragraph{Evaluation metrics.}

Referring from evaluation methodologies established in prior research on audio generation \cite{audioldm-liu2023audioldm}, our study employs frechet distance (FD), inception score (IS), and kullback-leibler (KL) divergence to assess the quality of generated musical audio outputs. 
(\textbf{i}) FD metric gauges audio quality by leveraging an audio embedding model to quantify the similarity between the embedding spaces of the generated and target audio. 
In this investigation, we employ two standard audio embedding models: VGGish \cite{hershey2017cnn} and PANN \cite{pann}, denoting the resulting distance as $FD_{vgg}$ and $FD_{pann}$, respectively. 
(\textbf{ii}) IS serves to measure the diversity and the quality of the full set of audio outputs.
(\textbf{iii}) KL divergence is measured on individual pairs of generated and ground-truth audio samples and averaged. 
We use the \texttt{audioldm\_eval} library\footnote{https://github.com/haoheliu/audioldm\_eval} to execute the evaluation of all aforementioned metrics. The comparative assessment involves juxtaposing the ground truth audio sourced from the Audiostock 1000-track test set with the 1000 tracks of music generated by each system, grounded in their corresponding textual descriptions.

\paragraph{Implementation details.}

For convenience, we opt to directly utilize the existing pre-trained music diffusion models MusicLDM\footnote{https://github.com/RetroCirce/MusicLDM} as the teacher model while maintaining fixed network parameters throughout the consistency distillation process.
The model structure of the consistency model is the same as the teacher diffusion model and is initialized with the teacher's model parameters. 
AdamW optimizer with a learning rate of 1e-5 is adopted to train MusicCM. The EMA rate used in our experiments is 0.95. 
We trained all MusicLDM modules with music clips of 10.24 seconds at a 16khz sampling rate. In both the VAE and diffusion model, music clips are represented as mel-spectrograms with $T=1024$ frames and $F=128$ mel-bins. VAE utilizes a downsampling rate of $P=8$ and a latent dimension of $C=16$. The architecture of MusicCM's latent diffusion model follows that of MusicLDM.
We use the ViT-B AudioMAE from repository\footnote{https://github.com/facebookresearch/AudioMAE} as a discriminator feature network by default. 
The distillation process incorporates a weighting factor of $\lambda=3$ consistently across all experiments according to prior experiments.

\subsection{Time Efficiency}

We quantify the inference time entailed in the process of text-to-music synthesis using our proposed \texttt{MusicCM}, conducting a comparative analysis against both the baseline method and other open-source models. The outcomes of this comparative evaluation are delineated in Table \ref{tab:1}. The findings reveal a noteworthy efficiency advantage for our proposed approach, as it necessitates a mere 1-4 steps for inference, rendering it considerably swifter compared to the baseline method requiring 50 DDIM steps.
It is imperative to emphasize that the inference cost encompasses not only iterative denoising but also other components such as text feature encoding and latent code decoding. 
More encouragingly, our method's efficacy is further underscored by its capacity to synthesize longer music sequences with diminished memory consumption and temporal overhead. To illustrate, on a single NVIDIA A100 GPU, our approach can generate a one-minute music clip in merely 1.2 seconds. This comparative analysis underscores the pronounced efficiency of our proposed approach.

\subsection{Quantitative Comparison to State-of-the-Art}

\paragraph{Generation quality.}
We present the FD, IS, and KL results in comparison with baseline models, as summarized in Table \ref{tab:1}. The results of Riffusion and MuBERT are reported from the paper \cite{chen2023musicldm}. 
Upon comparative analysis, our findings indicate that \texttt{MusicCM} (1-step) achieves a level of performance on par with Riffusion and MuBERT. This parity is attributed to our contributions in optimizing distillation datasets and employing a robust baseline model. Furthermore, \texttt{MusicCM} (4-steps) exhibits superior performance compared to the benchmark AudioLDM across nearly all metrics, with only marginal performance degradation compared to MuiscLDM. We attribute these advantages to the efficacy of score distillation incorporated in our methodology.

\begin{table}[t]
\caption{\textbf{Evaluation results for measuring the text-audio relevance and novelty (plagiarism)}. The results of \texttt{MusicCM} are calculated for four inference steps. }
\label{tab:2}
\resizebox{0.48\textwidth}{!}{
\begin{tabular}{lccc}
\toprule
 & \multicolumn{1}{c}{\textbf{Relevance}} & \multicolumn{2}{c}{\textbf{Novelty and Plagiarism Risk}} \\ \midrule
 & \multicolumn{1}{c}{TA Similarity$\uparrow$} & $SIM_{AA}@90\downarrow$ & $SIM_{AA}@95\downarrow$ \\ \midrule
Test Set & \multicolumn{1}{c}{0.325} & --- & ---  \\
Retrieval Max & \multicolumn{1}{c}{0.423} & --- & ---  \\ \midrule 
MuBERT & \multicolumn{1}{c}{0.131} & 0.107 & 0  \\
MusicLDM & \multicolumn{1}{c}{{0.281}} & 0.430 & 0.047\\
 \texttt{MusiCM} &  \multicolumn{1}{c}{{0.275}} & 0.330 & 0.021 \\ \bottomrule 
\end{tabular}}
\end{table}

\paragraph{Text-audio relevance, novelty, and plagiarism.} 
We also conduct an evaluation using two metric groups, as stipulated by \cite{chen2023musicldm}. The assessment aims to gauge text-audio relevance, novelty, and potential plagiarism risks inherent in diverse models.
(\textbf{i}) Text-audio similarity quantifies the relevance between the text and the audio in a common embedding space. 
(\textbf{ii}) To ascertain the degree to which models exhibit direct replication of samples from the training set, we initially compute the dot products between the audio embedding of each generated music output and all audio embeddings within the Audiostock training set. Subsequently, we identify the maximum similarity among the nearest neighbors in the training set. The nearest-neighbor audio similarity ratio is then calculated as the fraction of generated outputs for which the similarity of the nearest-neighbors exceeds a specified threshold. This ratio is denoted as $SIM_{AA}@90$ when the threshold is set to 0.9 and $SIM_{AA}@95$ with a threshold of 0.95. 
Table \ref{tab:2} delineates the mean values of text-audio similarity and nearest-neighbor audio similarity ratios, employing two distinct thresholds, for the 1000 tracks within the Audiostock test set and the generated music outputs from MuBERT, MusicLDM, and \texttt{MusicCM}.  
Two reference points are incorporated for the assessment of text-audio similarity: “Test Set” and “Retrieval Max”.
The outcomes reveal that the distilled $\texttt{MusicCM}$ exhibits a text-music alignment capability akin to the baseline while notably enhancing novelty and mitigating the risk of plagiarism.

\subsection{Model Analysis}

We conduct an ablation study on several choices in Table \ref{tab:3}. We will discuss each part in the following.

\begin{figure*}[t]
   \centering
   \includegraphics[width=1\linewidth]{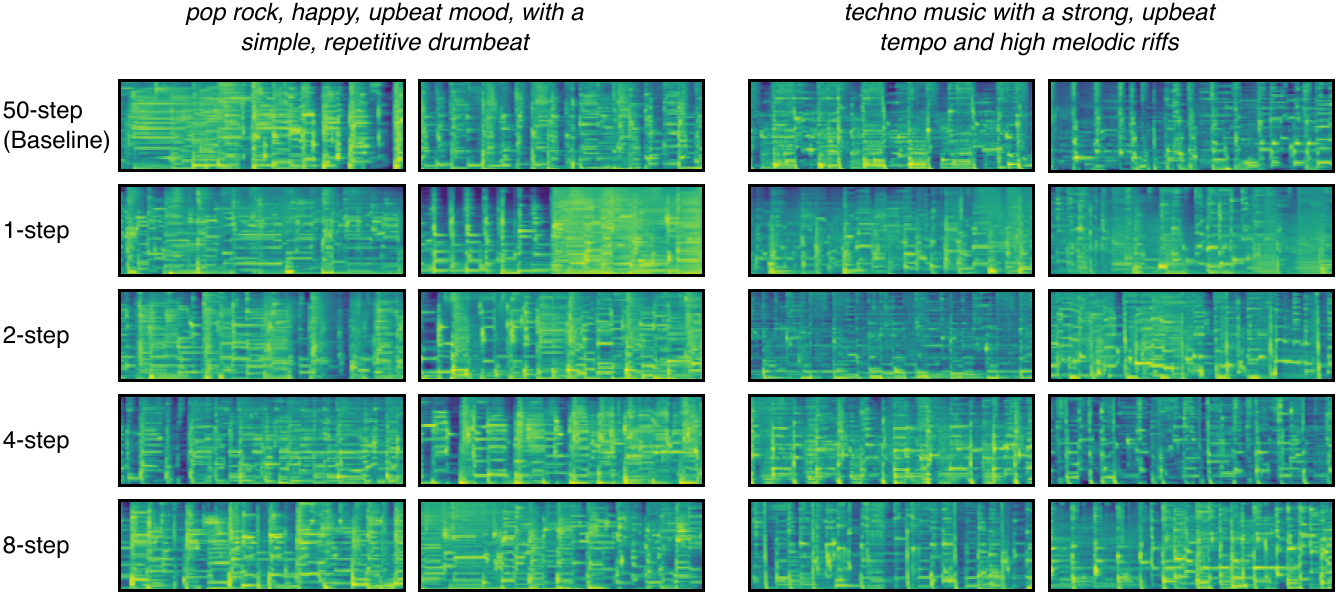}
   \caption{\textbf{Qualitative visualization results under different inference steps.} Larger steps generally yield better quality and time continuity of music. Importantly, our \texttt{MusicCM} can produce plausible results with fewer steps or even only one step. 
   }
   \label{fig:3}
\end{figure*}

\begin{table}[t]
\caption{\textbf{Ablation study.} We report FID, Is, KL, and text-audio similarity metrics in different settings. The results are calculated for four inference steps. }
\label{tab:3}
\small
\centering 
\begin{tabular}{lcccc}
\toprule
\textbf{Discriminator} & \textbf{FD}$_{pann}$ $\downarrow$ & \textbf{IS} $\uparrow$ & \textbf{KL} $\downarrow$ & \textbf{TA similarity} $\uparrow$\\
\midrule
\texttt{MusicCM} & 27.13 &1.78 &3.88 &0.275\\
\emph{w.o.} $L_{adv}$& 28.65 &1.73&3.95 &0.271 \\
\emph{w.} CLAP &27.55 &1.74 &3.92 &0.295 \\
\bottomrule
\end{tabular}
\end{table}

\paragraph{Adversarial loss term.}
In accordance with the findings presented in line 2, it becomes evident that adversarial losses play a crucial role. While the distillation loss in isolation demonstrates a certain level of effectiveness, its amalgamation with the adversarial loss yields a discernible enhancement in the obtained results. Notably, in our preceding human testing endeavors, a more intuitive comprehension is achieved. Particularly, when generating content with elaborate and extensive prompts, the incorporation of adversarial loss yields a perceptibly clearer and higher-quality output of musical clips.

\paragraph{Discriminator feature networks.} 

Recent insights suggest that ViT trained with CLIP or MAE exhibits particular aptitude in assessing the performance of generative models. Concurrently, these models also demonstrate efficacy when employed as discriminator feature networks. As delineated in line 3, we subsequently substitute AudioMAE with CLAP, resulting in an optimization in the aspect of text-audio similarity but a concomitant loss in preserving essential audio characteristics. Upon holistic consideration of the overall outcomes, AudioMAE emerges as the preferred choice. Nonetheless, it is acknowledged that the pursuit of more advanced feature networks for music perception remains a direction for future exploration.

\begin{table}[t]
\caption{\textbf{Subjective listening test} to evaluate different aspects including quality, relevance, and musicality for model outputs. }
\centering 
\small 
\label{tab:hm}
\begin{tabular}{lcccc}
\toprule
 &\textbf{Time}& \textbf{Quality} $\uparrow$ & \textbf{Relevance} $\uparrow$ & \textbf{Musicality} $\uparrow$ \\  \midrule 
MusicLDM &10s & 0.341&0.283& 0.541 \\ 
\texttt{MusicCM} &10s & 0.358 &0.267&  0.517     \\
MusicLDM &60s & 0.292 & 0.275 & 0.516 \\ 
\texttt{MusicCM} &60s & 0.350 & 0.291 & 0.525   \\
\bottomrule 
\end{tabular}
\end{table}

\subsection{Subjective Listening Test}
We additionally undertake a subjective listening test involving MusicLDM and \texttt{MusicCM} (4-steps) across varying durations of music clips to conduct a more nuanced assessment of the perceived auditory experience elicited by the generated music, as illustrated in \ref{tab:hm}. In this specific evaluation, we enlist the participation of 8 subjects who are instructed to listen to two distinct groups of generations randomly selected from the test set. Subsequently, they are tasked with determining whether the music is generated by the model or created by a human. Each group comprises 15 generations from both models, accompanied by their respective text descriptions.
The subjects are then prompted to rate the music based on three criteria: (\textbf{i}) Quality, pertaining to the sound quality of the generated samples; (\textbf{ii}) Relevance, assessing the alignment of the music with the provided text; and (\textbf{iii}) Musicality, gauging the overall aesthetic appeal of the music.
Our observations indicate the following: (\textbf{i}) For 10-second music generation, the samples generated by the \texttt{MusicCM} strategy exhibit superior music quality compared to those of the original MusicLDM, while remaining comparable in terms of relevance, thereby reinforcing the earlier analysis; (\textbf{ii}) In the generation of longer music sequences, our \texttt{MusicCM} approach demonstrates a significant superiority in terms of music quality. This enhancement is attributed to the incorporation of a shared restricted diffusion process.

\subsection{Qualitative Results}
In addition to our quantitative analyses outlined above, we present qualitative results pertaining to varying inference steps in Figure \ref{fig:3}. The findings elucidate that when the sampling step is restricted, as exemplified by step = 1, the generated mel-spectrogram exhibits perceptible blurriness, manifesting inaccuracies in detailing and an inability to preserve the temporal structure of objects. With an increment in the number of iteration steps, there is a discernible enhancement in the quality of the mel-spectrograms, concomitant with an improved preservation of temporal-frequency structures.
For instance, with 4-6 steps, outcomes comparable to DDIM's 50 steps can be attained, thereby substantively reducing sampling steps and concurrently enhancing generation speed. The outcomes obtained in the domain of text-to-music generation underscore the efficacy of the proposed \texttt{MusicCM}, demonstrating a commendable equilibrium between quality and speed.

\section{Limitations}
In this section, we outline the recognized limitations of our study, serving as a roadmap for future improvements.
Firstly, our \texttt{MusicCM} relies on a strong teacher model as the distillation target. It is hard to combine teachers from different domains or different music sampling ratios. Secondly, the consistency distillation process requires fine-tuning the model. While consistency distillation only requires a small number of training steps, it may lead to unsatisfactory results when the training data for the teacher model is unavailable. Finally, we also intend to investigate the optimization of different discriminator feature networks during adversarial training.

\section{Conclusion}
This study introduces music consistency models, presenting a universal approach for distilling a pre-trained diffusion model into a swift, few-step music generation model. The amalgamation of an adversarial and a distillation objective is employed for the distillation of robust baseline models. Empirical results substantiate the efficacy of our approach, illustrating the attainment of high-fidelity music synthesis within merely four steps. This outcome underscores its potential for real-time synthesis, a notable advancement over preceding methods necessitating approximately 50 DDIM steps. Our model facilitates the generation of high-quality music in a single step, thereby opening new avenues for real-time generation utilizing foundational models.

\bibliographystyle{named}
\bibliography{ijcai24}

\end{document}